\documentclass[12pt]{iopart}
\begin{document}
\title{The Nonperturbative Propagator and Vertex in Massless Quenched 
$QED_d$}

\author{A. Bashir$^{1,2}$ and R. Delbourgo$^2$} 

\address{$^1$ Instituto de F{\'\i}sica y Matem\'aticas,
Universidad Michoacana de San Nicol\'as de Hidalgo, Apartado Postal
2-82, Morelia, Michoac\'an 58040, M\'exico}

\address{$^2$ School of Mathematics and Physics, University of 
Tasmania, Hobart 7001, Australia} 

\ead{adnan@itzel.ifm.umich.mx, Bob.Delbourgo@utas.edu.au}

\begin {abstract}
It is well known how multiplicative renormalizability of the fermion
propagator, through its Schwinger-Dyson equation, imposes restrictions on the 
3-point fermion-boson vertex in massless quenched quantum electrodynamics in 
4-dimensions (QED$_4$). Moreover, perturbation theory serves as an excellent 
guide for possible nonperturbative constructions of Green functions.
 
We extend these ideas to arbitrary dimensions $d$. The constraint of 
multiplicative renormalizability of the fermion propagator is generalized to 
a Landau-Khalatnikov-Fradkin transformation law in $d$-dimensions and it 
naturally leads to a constraint on the fermion-boson vertex. We verify that 
this constraint is satisfied in perturbation theory at the one loop level in 
3-dimensions. Based upon one loop perturbative calculation of the vertex, we 
find additional restrictions on its possible nonperturbative forms in 
arbitrary dimensions.
\end{abstract}

\submitto{\JPA}
pacs{11.10.Kk, 11.15.Tk, 12.20.-m}

\maketitle

\section{Introduction}

The behaviour of the fermion propagator and the fermion-boson vertex 
in quantum electrodynamics (QED), along with higher Green functions, is 
dictated by the corresponding Schwinger-Dyson equations (SDE). However,
one expects that the gauge theory conditions such as the 
Ward-Green-Takahashi identity (WGTI),~\cite{WGTI} \footnote{It was also
given by Fradkin, see the corresponding reference in~\cite{LKFT}.}, 
and the Landau-Khalatnikov-Fradkin 
transformations (LKFT),~\cite{LKFT}, (derived also by Johnson and Zumino
through functional methods~\cite{zumino}),
should impose simplifying constraints on the equations that these
functions satisfy. As these conditions are nonperturbative in nature,
one might hope to arrive at a better understanding of the propagator as well 
as the vertex to an arbitrary order in perturbation theory. The WGTI 
is relatively straightforward to implement and its constraints easier 
to understand. In contrast, extracting useful information from the LKFT 
has proved a harder nut to crack except only in a small number of special 
cases and even then success has been achieved to a limited extent. 

The WGTI for the vertex determines what is often called its 
longitudinal part,~\cite{BC}. In the quenched version of 4-dimensional QED 
(QED$_4$), it has long been known that the remaining transverse part plays a 
crucial role in ensuring the correct LKFT or, in other words, the 
multiplicative renormalizability of the fermion propagator,~\cite{King,Brown}. 
This consideration was taken into account in~\cite{CP} and ~\cite{BP1} to 
propose a transverse vertex. These {\em ansatze} ensure that the fermion 
propagator is multiplicatively renormalizable. However, the construction 
involves the assumption that the anomalous dimension $\gamma$ is zero in the 
Landau gauge; crucially this is not the general solution, and disagrees 
with perturbation theory beyond one loop,~\cite{Ross,Tarasov}. This was noted 
and fixed in~\cite{BKP1}; there it was shown that a fermion propagator which 
obeys its LKFT at every order of perturbation theory satisfies a simple 
equation in momentum space. Moreover, it governs a transverse part of the 
fermion-boson vertex which can be reduced just to one unknown function. 
In the same article, this function was evaluated at the one loop level in the 
limit when momentum in one of the fermion legs is much greater than in the
other, imposing further constraints on the possible forms of the
vertex. However, this effort was restricted only to four dimensions.
In this paper, we investigate how these arguments could be extended
to arbitrary dimensions for the two and three point Green functions.

There exists some earlier works in that context ~\cite{BR,Dong}. 
In perturbation theory, these works use the restrictions imposed by
the two-point Green function at the one loop level to constrain the vertex. 
In the present article, we pose the question whether the fermion propagator 
could satisfy a simple integral equation in momentum space in $d$-dimensions 
just like the one in 4-dimensions proposed in~\cite{BKP1} such that
the gauge covariance is preserved not just to one loop but to all orders in 
perturbation theory. It appears that the answer is in the affirmative. 
We propose such an equation, which happily reduces to the one in~\cite{BKP1}
for the special case of $d=4$. Moreover, we verify that in 3-dimensions,
it gives us the correct gauge covariance of the fermion propagator up to 
three loops. We also argue that the terms of the type $(\alpha \xi)^n$ in 
the fermion propagator governed by this equation will be correct in arbitrary 
dimensions to {\em every order in perturbation theory}.

The SDE for the fermion propagator ties it to the fermion-boson vertex and
imposes a direct constraint on the vertex.  We verify that such a constraint 
is indeed satisfied in 3-dimensions at the one loop level for the vertex. 
In addition to this 
restriction, we know that perturbation theory plays a vital role in providing 
us with a guide to which every nonperturbative {\em ansatz} of the vertex 
reduce in the region of weak coupling. Whereas constructing such a vertex 
is highly non-trivial for all values of the fermion momenta involved, except 
in lower dimensions~\cite{BR1}, it is more likely to be achieved for
some asymptotic limits. We will largely focus on the limit when momentum in 
one of the fermion legs is much greater than the one in the other n the
Euclidean regime. Knowing the one loop vertex in this limit is made possible 
by the recent perturbative calculation of the transverse vertex by Davydychev 
{\em et. al.},~\cite{davydychev}, carried out in an arbitrary covariant 
gauge in arbitrary dimensions. In the said limit, we find a neat relation 
between the transverse vertex and the fermion propagator, reminiscent of 
the one which exists between the longitudinal vertex and the fermion 
propagator. As a verification of our result, we see that for the particular 
cases of $d=3$ and $d=4$, we recuperate the results known in literature.

\section{The massless fermion propagator}

The SDE for the unrenormalized fermion propagator $S(p)$, in quenched QED, 
with a bare coupling $e$ is given by~:
\begin{equation}
   S^{-1}(p)={\not \! p}+ie^2\int \frac{d^dk}{(2\pi)^d}\,
    \gamma^{\mu}\,S(k)\, \Gamma^{\nu}(k,p)\,\Delta_{\mu\nu}^{0}
    (q) \; , \label{eq:SDFP}
\end{equation}
where  $q=k-p$ and $d$ is the dimension of space-time. For massless fermions, 
$S(p)$ can be expressed in terms of a single Lorentz scalar function, $F(p^2)$, 
associated with the wavefunction renormalization, so that 
$S(p)= F(p^2)/\not\!p$; the bare propagator has $F=1$. 
The photon propagator remains unrenormalized in quenched QED and is
given by $ \Delta_{\mu\nu}^0(q)= \left(
    g_{\mu\nu}+(\xi-1) {q_{\mu}q_{\nu}}/{q^2} \right)/q^2 $,
where $\xi$ 
is the standard covariant gauge parameter. $\Gamma^{\mu}(k,p)$ is the full 
fermion-boson vertex for which we must make an {\em ansatz}. Ball and 
Chiu~\cite{BC} considered the vertex as a sum 
of longitudinal and transverse components. i.e., $  \Gamma^{\mu}(k,p)
= \Gamma^{\mu}_{L}(k,p)+\Gamma^{\mu}_{T}(k,p)$,
where $\Gamma^{\mu}_{T}(k,p)$ is defined by $
q_{\mu} \Gamma^{\mu}_{T}(k,p)=0$.  
To satisfy Eq.~(\ref{eq:SDFP}) in a manner free of kinematic singularities,
which in turn ensures the Ward identity is fulfilled, we have a possible
solution (following Ball and Chiu)~:
\begin{equation}
\Gamma^{\mu}_{L}(k,p) = {\cal A}(k^2,p^2) \gamma^{\mu}
              + {\cal B}(k^2,p^2) (\not \! k + \not \! p)
               (k+p)^{\mu} \; ,  \label{eq:BCV}
\end{equation}
 where
\begin{eqnarray}
\nonumber
  {\cal A}(k^2,p^2)&=&\frac{1}{2}\ \left( \frac{1}{F(k^2)}\ +
  \frac{1}{F(p^2)}\  \right) \; , \\
  {\cal B}(k^2,p^2)&=&\frac{1}{2}\ \left(
  \frac{1}{F(k^2)} - \frac{1}{F(p^2)}\ \right)
\frac{1}{k^2 - p^2} \;,    \label{eq:BCAB}
\end{eqnarray}
and $\Gamma^{\mu}_{T}(p,p)=0$.
Ball and Chiu,~\cite{BC}, demonstrated that a set of 8 vectors 
$T_{i}^{\mu}(k,p)$ formed a general basis 
(slightly modified later by K{\i}z{\i}lers\"{u} {\em et. al.},~\cite{Ayse})
for the transverse part. However, for the 
massless case, only four of the basis vectors contribute to the fermion 
propagator equation and they are the same in both the schemes. Therefore, 
we can write~:
\vspace {2mm}
\begin{eqnarray}
   \Gamma^{\mu}_{T}(k,p)= \sum_{i=2,3,6,8} \tau_{i}(k^2,p^2,q^2)\ T_{i}^
     {\mu}(k,p)  \;.  \label{eq:basis}
\end{eqnarray}
The condition $\Gamma^{\mu}_{T}(p,p)=0$ is then satisfied
provided that in the limit $k\rightarrow p $, the $\tau_{i}(p^2,p^2,0)$
are finite. One can then define the Minkowski space basis in the massless 
case to be~\cite{BC,Ayse}~:
\begin{eqnarray}
 \nonumber  \\ T_{2}^{\mu}(k,p)&=&  (p^{\mu} k\cdot q - k^{\mu}  p\cdot q) 
( \not \! k +\not \! p)    \nonumber  \\
 \nonumber   T_{3}^{\mu}(k,p)&=& q^{2} \gamma^{\mu}-q^{\mu} \not \! q     \\
 \nonumber   T_{6}^{\mu}(k,p)&=& \gamma^{\mu} (k^{2} - p^{2})
             -( k+p)^{\mu} ( \not \! k -\not \! p )                    \\
  T_{8}^{\mu}(k,p)&=&  \gamma^{\mu} \sigma_{\lambda\nu} k^{\lambda} p^{\nu}  
               - k^{\mu} \not \! p + p^{\mu} \not \! k   
\; . \label{eq:basis1} \\ \nonumber 
\end{eqnarray}
On multiplying Eq.~(\ref{eq:SDFP}) by  $\not \! p$, taking
the trace, making use of 
Eqs.~(\ref{eq:BCV},\ref{eq:BCAB},\ref{eq:basis},\ref{eq:basis1}), and 
Wick rotating to Euclidean 
space, we obtain~:
\begin{eqnarray}
\nonumber
 \frac{1}{F(p^2)}\,=\; 1\;-
           &&  \frac{e^2}{p^2} \int
               \frac{d^dk}{(2 \pi)^d} \,
               \frac{F(k^2)}{k^2q^2}\
                \\   \nonumber   \
           & &  \Bigg\{ \vspace{5 mm} {\cal A}(k^2,p^2)
               \frac{1}{q^2}
             \left[ -2 \Delta^2 + (1-d) q^2 k\cdot  p \right]
              \\  \nonumber
           & &+  {\cal B}(k^2,p^2) \frac{1}{q^2}
            \left[ -2 (k^2+p^2) \Delta^2 
            \right]  \\  \nonumber
           & &- \frac{\xi} {F(p^2)}\ \frac{p^2}{q^2}
               \ (k^2-k\cdot p)  \\ \nonumber
           & &+\tau_{2}(k^2,p^2,q^2)
               \left[- (k^2+p^2) \Delta^2  \right]   \\ \nonumber
            \nonumber
           & &+\tau_{3}(k^2,p^2,q^2)
               \left[ - 2 \Delta^2 + (1-d) q^2 k\cdot p  \right]
             \\  \nonumber
           & &+\tau_{6}(k^2,p^2,q^2)
               \left[ (1-d) (k^2-p^2) k\cdot p  \right] \\
           & &+\tau_{8}(k^2,p^2,q^2)
               \left[ (d-2) \Delta^2 \right] \Bigg\} \;, \label{eq:FEX}
\end{eqnarray}
where $\Delta^2= (k \cdot p)^2 - k^2 p^2 $.
If we knew what the $\tau_i(k^2,p^2,q^2)$ were, we could solve this equation 
for the fermion propagator. In practice, it is obviously not an easy task, 
since these transverse functions are only known in perturbation theory or are 
connected to higher order Green functions by gauge identities and SDEs.
It is conceivable that only a small part of the right hand side of 
Eq.~(\ref{eq:FEX}) is sufficient to yield the main characteristics of the 
full propagator. We look at this possibility in the next subsection, but 
first we note that  $\cal B$ is of order $e^2$ like $\tau_i$ and both can only
contribute to $F$ at least to order $e^4$. The dominant term which is at least
of order $e^2$ in $F$ is associated with the $\xi$-piece.

\subsection{The $(\alpha \xi)^n$ terms}

Bashir and Pennington,~\cite{BP1}, noted that if one assumes the anomalous
dimension to vanish in the Landau gauge in the quenched case, the third term 
on the right hand side of Eq.~(\ref{eq:FEX}) gives the total answer for the 
fermion propagator in 4-dimensions. To begin with, let us so truncate Eq. (6) 
for arbitrary dimensions as well to see where this gets us. We would then deduce
\begin{eqnarray}
 \frac{1}{F(p^2)} &=& 1\;+ 
             \xi e^2 \int
         \frac{d^dk}{(2 \pi)^d} \,
         \frac{F(k^2)}{F(p^2)} \;\frac{k \cdot q}{k^2 q^4}\;. \label{eq:FEX1}
\end{eqnarray}
Thus in 4-dimensions, on carrying out the angular integration, one can
immediately see that the resulting equation has the 
solution $F(p^2) = (p^2/\Lambda^2)^{\nu}$,
Here $\nu=a \xi$, $a=\alpha/4 \pi$ and $\alpha= e^2/ 4 \pi$. The solution
correctly embodies the leading gauge terms which are proportional to 
$\alpha \xi$, $(\alpha \xi)^2$, $(\alpha \xi)^3$, etc. as is obvious in the
expansion  $F(p^2) = 1 + \nu \; {\rm ln} {p^2}/{\Lambda^2}
+ ({\nu^2}/{2!}) {\rm ln}^2 {p^2}/{\Lambda^2} + \cdots$.
What happens in arbitrary dimensions? Let us assume that Eq.~(\ref{eq:FEX1}) 
is correct for arbitrary dimensions. (Note that $e^2$ or $\alpha$ carry mass 
dimensions in arbitrary $d$.) In the following, we shall see what support we 
get from perturbation theory in favour of this assumption~:
\begin{itemize}

\item Let us first obtain the fermion propagator at order $\alpha$. To do so, 
we can set $F=1$ in Eq. (6) or ~(\ref{eq:FEX1}). A $d$-dimensional integration
can then be carried out and, on setting $d=4- 2 \epsilon$, we obtain~:
\begin{eqnarray}
F(\!p^2\!) \!=\! 1\! -\! \frac{e^2 \xi}{(4 \pi)^{2\!-\!\epsilon}} \; 
\frac{\Gamma(2\!-\!\epsilon) \Gamma(1\!-\!\epsilon) \Gamma(\epsilon)}{
\Gamma(2\! - \!2 \epsilon)} (p^2)^{-\epsilon}, 
\label{oneloopfp}
\end{eqnarray}
which is the {\em exact} one loop result in arbitrary dimensions,
~\cite{davydychev}.
At the one loop level, dimensional integration shows that the ${\cal A}$-term 
is identically zero (see appendix). In fact, interestingly, the ${\cal A}$-term
vanishes at all orders in perturbation theory. We again leave the details to 
the appendix. {\em This means that the ${\cal A}$-term in Eq.~(\ref{eq:FEX}) 
can be safely discarded once and for all. This is one of the significant 
new results of our article and it is completely independent of any 
truncations of Eq. (6)}

\item  Now that we know the one loop result, we can get the higher powers of
$(\alpha\xi)$ simply by successive substitution. Thus putting the above
answer back in Eq.~(\ref{eq:FEX1}) we get the two loop result. A simple 
exercise gives~:
\begin{eqnarray}
\nonumber F(p^2\!)\!\! &=&\! 1 - \frac{e^2 \xi}{(4 \pi)^{2-\epsilon}} \; 
\frac{\Gamma(2\!-\!\epsilon) \Gamma(1\!-\!\epsilon) \Gamma(\epsilon)}{
\Gamma(2 - 2 \epsilon)}\left( p^2 \right)^{-\epsilon}    \\
&-&\!\!\frac{e^4 \xi^2}{(4 \pi)^{4-2 \epsilon}}
\frac{\Gamma(2\!-\!\epsilon) \Gamma(1\!-\!\epsilon) \Gamma(- \epsilon) 
\Gamma(2 \epsilon)}{
\Gamma(2\! - \!3 \epsilon)} (p^2)^{-2 \epsilon}  \nonumber \\
\end{eqnarray}
which encapsulates the $(\alpha \xi)^2$ term in arbitrary 
dimensions. A complete calculation of the two loop fermion
propagator can be found in~\cite{Tarasov1}.
We can keep repeating the procedure and we shall get exact leading gauge
terms at all orders, i.e., terms of the type $\alpha \xi$, $(\alpha \xi)^2$,
$(\alpha \xi)^3$, $(\alpha \xi)^4$, and so on. Each time, the integral 
involved is similar to the one previously evaluated. Therefore, it is rather
easy to know the coefficient of the term $(\alpha \xi)^n$ for arbitrarily
large values of $n$.

\end{itemize}
One should note that although the above procedure accounts for all the terms
of the type $(\alpha \xi)^n$ at an arbitrary order of approximation, it does 
not incorporate non-leading gauge terms, e.g., the ones proportional 
to $\alpha^2$, $\alpha^3 \xi$, etc. Can a simple extension of 
Eq.~(\ref{eq:FEX1}) achieve this? We attempt to answer this question in the
next subsection.

\subsection{The complete propagator}

It was pointed out in~\cite{BKP1} that in 4-dimensions it is rather 
simple to generalize Eq.~(\ref{eq:FEX1}) to incorporate the non-leading
gauge terms. Therefore,~\cite{BKP1} demonstrates that
\begin{eqnarray}
 \frac{1}{F(p^2)} = 1 + 
           (4 \pi)^2 \gamma \int
               \frac{d^4k}{(2 \pi)^4} \,
               \frac{F(k^2)}{F(p^2)} \;   \frac{k \cdot q}{k^2 q^4}
\;,
\end{eqnarray}
where $\gamma\,=\,\gamma_0 +\,\alpha\xi/{4\pi}$ ($\gamma_0$ is gauge 
independent and, in perturbation theory, is of ${\cal O}(\alpha^2)$), yields 
the correct gauge covariance of the fermion propagator in four dimensions
in perfect agreement with the rules of the LKFT,~\cite{LKFT,zumino}. 

The question we raised was whether a similar equation also yields the 
correct covariance properties in arbitrary dimensions. As $\gamma$ becomes 
dimensionful in  arbitrary $d$, a simple extension of this equation might
read~:
\begin{eqnarray}
 \frac{1}{F(p^2)} = 1 + 
            (4 \pi)^2  \int
               \frac{d^dk}{(2 \pi)^d} \, \gamma_d(k^2,p^2,q^2)
               \frac{F(k^2)}{F(p^2)} \;   \frac{k \cdot q}{k^2 q^4} \;
, \label{FPddim}   \nonumber \\
\end{eqnarray}
where the the function $\gamma_d(k^2,p^2,q^2)$ necessarily depends upon $d$. 
Therefore, we can write it as follows~:
\begin{eqnarray}
   \gamma_d(k^2,p^2,q^2) &=& \frac{\alpha \xi}{4 \pi} + 
\lambda_d(k^2,p^2,q^2) \;.
\label{gamma}
\end{eqnarray}    
(The functions $\gamma_d(k^2,p^2,q^2)$ and $\lambda_d(k^2,p^2,q^2)$ 
acquire mass dimensions $m^{(4-d)}$ like $\alpha$.) In 4-dimensions, 
$\gamma_d(k^2,p^2,q^2)$ is nothing but the anomalous dimension and according 
to reference~\cite{Tarasov} $\gamma_4=\gamma= a \xi - \frac{3}{2} \, 
a^2 + \frac{3}{2} \, a^3 +  {\cal O}(a^4)$. 
Therefore, we can identify
\begin{eqnarray}
    \lambda_4(k^2,p^2,q^2) &=& \lambda_4 = - \frac{3}{2} \, a^2 
+ \frac{3}{2} \, a^3 +  {\cal O}(a^4) \;.   \label{anom}
\end{eqnarray}
In the spirit of 4-dimensions, we seek a $\lambda_d(k^2,p^2,q^2)$ that is 
{\em independent of the covariant gauge parameter in arbitrary
dimensions d}. Thus we have squeezed all the information contained in the 
full fermion-boson vertex (projected onto the fermion propagator)
into $\lambda_d(k^2,p^2,q^2)$. Whereas knowing
the full vertex at every order of $\alpha$ and thus calculating the
fermion propagator is a formidable task, we look for some simple parametric 
form of the function $\lambda_d(k^2,p^2,q^2)$ which would be sufficient to 
capture the correct gauge covariance of the propagator at all orders in 
arbitrary dimensions. As an example, we put to test the following form~:
\begin{eqnarray}
\nonumber   \lambda_d(k^2,p^2,q^2) &=& \alpha^2 
\left[ \frac{\lambda_1^d}{(k^2)^{2-d/2}} 
+ \frac{\lambda^{\prime d}_1}{(p^2)^{2-d/2}}   \right] \\
&+& \alpha^3 \left[ \frac{\lambda_2^d}{(k^2)^{2(2-d/2)}} 
+ \frac{\lambda^{\prime d}_2}{(p^2)^{2(2-d/2)}}  \right] \nonumber \\
&+& {\cal O}(\alpha^4) \;, \label{lambda}
\end{eqnarray}
where we expect the constants $\lambda_i$ and $\lambda^{\prime}_i$ all 
to be independent of the gauge parameter $\xi$. [Note that we have avoided
including $q^2$ dependence in (29). Our guiding principle was to choose a form 
as simple as possible without spoiling the objective of the exercise. We could 
readily have introduced mixed terms of the type $1/(k^2p^2)^{2-d/2}, 
\Delta^{-2+d/2}$ or even terms incorporating $q^2$ to the appropriate power. 
It is perfectly conceivable that they can arise in more complicated 
topological contributions to the vertex function but we have strong
hints that they are absent in the asymptotic regime $k^2 >> p^2$ which
we shall be considering shortly.]
Eqs.~(\ref{FPddim},\ref{gamma},\ref{lambda}) form the main equations of this 
section. We point out that for $d$=4, the momentum dependence of the 
gamma function disappears as expected and we can identify, $\lambda_i^4$
and $\lambda_i^{\prime 4}$ by making a direct comparison with Eq.~(\ref{anom}).

Before we try to see if Eqs.~(\ref{FPddim},\ref{gamma},\ref{lambda}) give the 
propagator equation 
with correct properties in arbitrary dimensions, let us clarify its status.
This equation does not determine the values of $\lambda_i$ and 
$\lambda^{\prime}_i$ and such like. However, given such constants, one can 
determine the propagator in an arbitrary covariant gauge. This is similar to 
what the LKFT specify: given the fermion propagator in one particular gauge,
they determine what the propagator will be in an arbitrary covariant
gauge. However, LKFT are written in the coordinate space and the Fourier 
transforms involved can only be performed easily in certain very
limited number of special cases. On the other hand, solving Eq.~(\ref{FPddim})
is much simpler, as we have demonstrated early on.

We do not attempt here to prove the validity of 
Eqs.~(\ref{FPddim},\ref{gamma},\ref{lambda}) in all dimensions, but we
expect that some simple parametric form of $\lambda_d(k^2,p^2,q^2)$, like ours, 
is able to give the correct gauge dependence of the full fermion propagator. 
For now we only verify that this form does indeed yield the correct
gauge dependence of the fermion propagator for $d=3$ at least up to three 
loops in perturbation theory. 
A straightforward calculation reveals that if we select~:
\begin{eqnarray}
      \lambda_1^3  &=& \frac{3}{16} \left(\frac{7}{3} - \frac{\pi^2}{4} 
\right)  \;,     \nonumber \\
      \lambda^{\prime 3}_1 &=& 0 \;,   \nonumber
\end{eqnarray}
Eq.~(\ref{FPddim}) gives the following result until ${\cal O}(\alpha^3)$~:
\begin{eqnarray}
F(p^2) &=& 1 - \frac{\pi}{4} \frac{(\alpha \xi)}{p} + 
\frac{1}{4} \frac{(\alpha \xi)^2}{p^2}
- \frac{3}{4} \, \left(\frac{7}{3} - \frac{\pi^2}{4} \right) \, 
\frac{\alpha^2}{p^2} 
- \pi^2 \lambda^{\prime 3}_2 \frac{\alpha^3}{p^3} \;.
  \label{FPQED33} 
\end{eqnarray}
Comparing it with the findings in~\cite{BKP2} which gives the massless 
fermion propagator
to two loops in QED3, we confirm our truncation to this order. Moreover,
~\cite{B1} tells us that ${\cal O}(\alpha^3)$ term must be independent of
the gauge parameter. This is again in perfect agreement with our assumption 
that all the $\lambda_i$ and $\lambda^{\prime}_i$ (in this case 
$\lambda^{\prime 3}_2$) must be independent of the gauge parameter.

This example provides a verification that our proposal for the $\lambda_d$ 
function works for QED$_3$ up to three loops in addition to working for 
QED$_4$ at all orders. The $\lambda_i^3$ and $\lambda_i^{\prime 3}$ are
gauge independent quantities as we had claimed. Moreover, we do not get any 
terms proportional to $(\alpha \xi)^3$, $\alpha^3 \xi^2$ and $\alpha^3 \xi$, 
as already expected from the LKFT~\cite{B1}. It would be worth checking if 
this simple form of the function $\lambda_d(k^2,p^2,q^2)$ could yield the 
correct form of the fermion propagator in three dimensions at higher orders 
and in other dimensions but we do not address this issue in this paper. In 
the next section, we discuss the constraints imposed by our proposal and 
perturbation theory for the fermion propagator on the 3-point fermion-boson 
vertex.

\section{Asymptotic Constraints}

\subsection{Constraints from the SDE on $S(p)$}

After demonstrating with examples that 
Eqs.~(\ref{FPddim},\ref{gamma},\ref{lambda}) may account for the fermion 
propagator, should separate the rest of the terms in Eq.~(\ref{eq:FEX}),
which start off at order $\alpha^2$. These ought then to conspire against 
each other to give zero in the asymptotic regime at the very least, viz. 
\begin{eqnarray}
\nonumber
           && \int   \frac{d^dk}{(2 \pi)^d} \,
               \frac{F(k^2)}{k^2q^2}\
                \\   \nonumber   \
           & &  \Bigg\{   {\cal B}(k^2,p^2) \frac{1}{q^2}
            \left[ -2 \Delta^2 (k^2+p^2)
            \right]  \nonumber \\  \nonumber
           & &+ \frac{\lambda_d(k^2,p^2,q^2)} {\alpha F(p^2)}\ \frac{p^2}{q^2}
               \ (k^2-k\cdot p)  \\ \nonumber
           & &+\tau_{2}(k^2,p^2,q^2)
               \left[- \Delta^2  (k^2+p^2) \right]   \\ \nonumber
            \nonumber
           & &+\tau_{3}(k^2,p^2,q^2)
               \left[ - 2 \Delta^2 + (1-d) q^2 k\cdot p  \right]
             \\  \nonumber
           & &+\tau_{6}(k^2,p^2,q^2)
               \left[ (1-d) (k^2-p^2) k\cdot p  \right] \\
           & &+\tau_{8}(k^2,p^2,q^2)
               \left[ (d-2) \Delta^2 \right] \Bigg\} =0 \;. \label{conv}
\end{eqnarray}
This equation naturally places a constraint on the choice of the transverse 
part of the fermion-boson vertex. Any {\em ansatz} for the transverse vertex 
must ensure that this condition is satisfied for $k^2>>p^2$ anyhow. 

In principle, the knowledge of the transverse vertex at the lowest
order in arbitrary dimensions and gauge,~\cite{davydychev}, allows us
to verify this condition at that level of approximation. But the 
expressions for the $\tau_i$ are complicated and lengthy, rendering
the exercise highly non-trivial. However, we have managed to check
that at least for $d=3$, this condition is indeed satisfied. In a previous
work, it was checked also for $d=4$,~\cite{BKP1}.

\subsection{Perturbation theory constraints}

We know that perturbation theory is the only calculational scheme which
incorporates the key features of a gauge field theory such as gauge
identities at every order of approximation naturally. Therefore,
we stand a better chance in retaining these essential properties if
every nonperturbative {\em ansatz} of the 3-point vertex in truncating
the tower of Schwinger-Dyson equations makes sure that it reduces to its
perturbative Feynman expansion when the coupling involved is weak.
Whereas for arbitrary momenta of the fermions and photon, this expansion
can be very complicated, blurring our view of its possible nonperturbative
forms, certain asymptotic regimes of momentum can turn out to be
a better guide in this respect. In general, the $\tau_i$ are explicit 
functions of the angle between $k$ and $p$. Only in certain asymptotic 
limits does this dependence disappear. One such limit is when momentum
$k^2$ in one of the fermion legs is much greater than the one in the other 
leg, namely $p^2$. 
A recent calculation of Davydychev {\em et. al.} of the $\tau_i$ at the one 
loop level in arbitrary dimensions  enables us to take this limit. That is 
what we do next. One such integral that is somewhat more difficult to evaluate
in this limit is,~\cite{davydychev1},
\begin{eqnarray}
&&  J(4-2 \epsilon) = \int d^d \omega \; 
\frac{1}{\omega^2 (k- \omega)^2 (p-\omega)^2} \nonumber \\
&=& - 
\frac{2 \pi^{2 - \epsilon} i^{1+2 \epsilon}}{(-k^2 p^2 q^2)^{\epsilon}}
\frac{\Gamma^2(1-\epsilon) \Gamma(\epsilon)}{\Gamma(2-2 \epsilon)}\nonumber
\\
&& \nonumber \Bigg[\frac{(k^2 p^2)^{\epsilon}}{(k^2+p^2-q^2)}
~_2F_1\left( 1,\frac{1}{2};\frac{3}{2}-\epsilon; 
\frac{4 \Delta^2}{(k^2+p^2-q^2)^2}  \right)  \\
&&+ \frac{(k^2 q^2)^{\epsilon}}{(k^2+q^2-p^2)}
~_2F_1\left( 1,\frac{1}{2};\frac{3}{2}-\epsilon; 
\frac{4 \Delta^2}{(k^2+q^2-p^2)^2}  \right)  \nonumber  \\
&&+ \frac{(q^2 p^2)^{\epsilon}}{(q^2+p^2-k^2)}
~_2F_1\left( 1,\frac{1}{2};\frac{3}{2}-\epsilon; 
\frac{4 \Delta^2}{(q^2+p^2-k^2)^2}  \right) \nonumber \\
&&- \frac{\pi \Gamma(2-2 \epsilon)}{\Gamma^2(1-\epsilon)} \; 
\left[ -4 \Delta^2 \right]^{-\frac{1}{2}+ \epsilon} \; \theta_{123} \Bigg] \;,
\end{eqnarray}
where
\begin{eqnarray}
   \theta_{123} = \theta(k^2+p^2-q^2) \theta(k^2+q^2-p^2) 
\theta(q^2+p^2-k^2) \;.\nonumber
\end{eqnarray}
In the Euclidean limit $k^2 >> p^2$, the leading term  would be independent of 
the angle between $k$ and $p$. Therefore, the most convenient way to proceed is 
to choose a particular angle for which it is easier to take the limit. 
In order to confirm our results, we calculate $J$ for two angles between $k$
and $p$, namely $0$ and $\pi/2$, and expectedly arrive at the same result~:
\begin{equation}
  J(4\!-\!2 \epsilon)\!=\!- i \pi^{2\!-\!\epsilon} 
 \frac{\Gamma^2(1\!-\!\epsilon)\Gamma(\epsilon)}{\Gamma(2\!-\!2\epsilon)
 \;k^2} \left( (p^2)^{-\epsilon}\!-\!(k^2)^{-\epsilon}\right). \nonumber 
\end{equation}
In 3 and 4-dimensions, this expression reduces to the already known 
results,~\cite{BKP1,BKP2}. Further, it can now be related to the 2-point 
integral,~\cite{davydychev1},~:
\begin{eqnarray}
K(k^2;4-2 \epsilon) &=&  \int d^d \omega \; 
\frac{1}{\omega^2 (k- \omega)^2 } 
= i \pi^{2 - \epsilon} 
\; \frac{\Gamma^2(1-\epsilon) \Gamma(\epsilon)}{\Gamma(2-2 \epsilon) } \;
 (k^2)^{-\epsilon} \;.
\end{eqnarray}
Therefore, the 3-point integral can be written as the 
difference of two 2-point integrals in the limit when $k^2 >> p^2$~:
\begin{eqnarray}
 J(4-2 \epsilon) = \frac{1}{k^2} \; \left[ K(k^2;4-2 \epsilon) - 
K(p^2;4-2 \epsilon) \right] \;.
\end{eqnarray}
It is interesting to note that the lightlike limit of the vertex function
exhibits precisely the same form, viz. $[K(k^2)-K(p^2)]/(k^2-p^2)$, 
as has been pointed out in \cite{DOT,RD03}.

We may now anticipate that in this limit, the 3-point vertex will bear
a simple relationship with the 2-point propagator. However, for the
actual evaluation of the transverse vertex or the $\tau_i$ in this limit, 
we have to evaluate $J$ to the next order of approximation. On doing so, we 
obtain the following expression for the $\tau_i$~:
\begin{eqnarray*}
 \hspace{-23mm}
\tau_2(k^2,p^2) = \frac{e^2}{(4 \pi)^{2-\epsilon}} 
\frac{\Gamma(2-\epsilon) 
\Gamma(1-\epsilon) \Gamma(\epsilon)}{\Gamma(2-2 \epsilon)} 
 \left[ \frac{-1+(2-\epsilon) \xi }{3-2 
\epsilon} \right] \;
\frac{1}{k^4} \; \left(  (p^2)^{-\epsilon} - (k^2)^{-\epsilon}   \right) 
\\ \nonumber \\
\hspace{-23mm} \tau_3(k^2,p^2) = \frac{e^2}{2(4 \pi)^{2-\epsilon}} 
\frac{\Gamma(2-\epsilon) 
\Gamma(1-\epsilon) \Gamma(\epsilon)}{\Gamma(2-2 \epsilon)} 
\left[ \frac{2(1-\epsilon)-(2-\epsilon) \xi }{3-2 
\epsilon} \right] \;
\frac{1}{k^2} \; \left(  (p^2)^{-\epsilon} - (k^2)^{-\epsilon}   \right) 
\\ \nonumber \\
\hspace{-23mm} \tau_6(k^2,p^2) = \frac{e^2}{2(4 \pi)^{2-\epsilon}} 
\frac{\Gamma(2-\epsilon) 
\Gamma(1-\epsilon) \Gamma(\epsilon)}{\Gamma(2-2 \epsilon)} 
\left[ 
\frac{2(1-\epsilon)-(1-\epsilon) \xi }{3-2 \epsilon} \right] \;
\frac{1}{k^2} \; \left(  (p^2)^{-\epsilon} - (k^2)^{-\epsilon}   \right) \\
\nonumber \\
\hspace{-23mm} \tau_8(k^2,p^2) = \frac{e^2}{(4 \pi)^{2-\epsilon}} 
\frac{ 
\Gamma^2(1-\epsilon) \Gamma(\epsilon)}{\Gamma(2-2 \epsilon)}
 \left[ 1 + \epsilon \; \xi  \right] \;
\frac{1}{k^2} \; \left(  (p^2)^{-\epsilon} - (k^2)^{-\epsilon}   \right) \;.
\end{eqnarray*}
Any {\em ansatz} for the transverse vertex must reproduce these results in 
the weak coupling regime and in the asymptotic limit of momenta considered.
As a check, these expressions nicely reproduce the results quoted 
in~\cite{BKP1,BKP2} for the special cases of $d=4$ and $d=3$ respectively.
The $\tau_i$ can now be written in terms of the fermion propagator function
$F$, Eq.~(\ref{oneloopfp}), just like the coefficients of the longitudinal 
vertex. In the asymptotic limit mentioned above, a straightforward
exercise yields~:
\begin{eqnarray}
 \tau_2(k^2,p^2) &=& \frac{1-(2-\epsilon) \xi}{3-2 \epsilon} \frac{1}{k^4}
\left[ \frac{1}{F(k^2)} - \frac{1}{F(p^2)} \right]     \\
\tau_3(k^2,p^2) &=& -
\frac{2(1-\epsilon)-(2-\epsilon) \xi}{2(3-2 \epsilon)}  \frac{1}{k^2}
\left[ \frac{1}{F(k^2)} - \frac{1}{F(p^2)} \right]  \nonumber   \\ \\
\tau_6(k^2,p^2) &=& -
\frac{(1-\epsilon) (2-\xi)}{2(3-2 \epsilon)}  \frac{1}{k^2}
\left[ \frac{1}{F(k^2)} - \frac{1}{F(p^2)} \right]  \\
   \tau_8(k^2,p^2) &=&- \frac{1}{1-\epsilon} \; \frac{1}{k^2} 
\left[ \frac{1}{F(k^2)} - \frac{1}{F(p^2)} \right] \;.
\end{eqnarray}
Thus the $\tau_i$ come out to be related to the fermion propagator
in a strikingly similar manner as the coefficients of the longitudinal
vertex are related to it. It will 
be intriguing if these relations remain intact at higher orders in 
perturbation theory. Taking into account the tensor structures $T_i^{\mu}$, 
one can readily see that $\tau_2(k^2,p^2)$ and $\tau_8(k^2,p^2)$ provide 
{\em subleading} contributions to the transverse vertex. Therefore,
the complete transverse vertex {\em in this limit} can be written as
\begin{eqnarray}
\hspace{-23mm}
   \Gamma_{T}^{\mu}(k,p) = \frac{e^2\xi}{2(4 \pi)^{2-\epsilon}} 
\frac{\Gamma(2-\epsilon) 
\Gamma(1-\epsilon) \Gamma(\epsilon)}{\Gamma(2-2 \epsilon)} 
\frac{1}{k^2} \; \left(  (p^2)^{-\epsilon} - 
(k^2)^{-\epsilon}   \right) 
\; \left[ -k^2 \gamma^{\mu} + k^{\mu} \not \! k  \right] \;, 
\end{eqnarray}
which is a result of a magic cancellation of some $d$-dependent terms
in the expressions for $\tau_3$ and $\tau_6$. One can readily verify that
in particular cases of 4 and 3 dimensions respectively, this expression 
reproduces wellknown results,~\cite{BKP1,BKP2}. In terms of the fermion 
propagator, this expression can be written as~:
\begin{eqnarray}
 \Gamma_{T}^{\mu}(k,p) = \frac{\xi}{2 k^2} \; 
\left[ \frac{1}{F(k^2)} - \frac{1}{F(p^2)} \right] \; 
\left[ k^2 \gamma^{\mu} - k^{\mu} \not \! k  \right] \;.
\end{eqnarray}
This is very much like the longitudinal vertex written in terms of the
fermion propagator, guided by the WGTI. It will be interesting to see if 
this simple relation survives the complications of higher orders of
perturbation theory. The knowledge of the transverse vertex for an arbitrary 
regime of fermion momenta in its nonperturbative form is a highly nontrivial 
task. So far, only for 3-dimensions has such a vertex been proposed 
\cite{BR1}. However, we believe that perturbation theory at higher orders 
along with gauge identities should supply better insight in arbitrary 
dimensions.

\section{Conclusions}

The nonperturbative fermion propagator and the fermion-boson vertex
are in principle determined from the corresponding Schwinger-Dyson equations, 
although in practice, solving these equations exactly is a prohibitively 
difficult task. However, covariance gauge identities impose strict constraints 
on how these Green functions are related to one another and how they evolve 
under a variation of gauge. Perturbation theory too is an important guide 
where the systematic scheme of approximation ensures that these identities are 
satisfied identically at every order of $\alpha$. Based upon these ingredients, 
we have analyzed the fermion propagator and the fermion-boson vertex in a 
nonperturbative fashion in arbitrary dimensions. We have proposed a simple 
equation for the fermion propagator written in the momentum space which 
seems a likely candidate to capture the correct gauge covariance of this Green 
function, at asymptotic values in any event. As it is written in the 
momentum-space, it is much easier to extract information compared with the 
coordinate space LKFT. Guided by this equation and perturbation theory, we 
have arrived at (asymptotic) constraints on the fermion-boson vertex. We hope 
that higher order perturbative calculations and the LKFT for the vertex itself
could help us further pin down the possible nonperturbative forms of 
these Green functions in arbitrary dimensions. All this is for the future.

\vspace*{4mm}

\noindent
{\bf Acknowledgments}

\vspace*{3mm}

We are grateful to Andrei Davydychev for helpful discussions and bringing
some useful references to our attention. We also thank the Australian
Research Council for financial support under Grant No. A00000780.
AB wishes to thank the School of Mathematics and Physics, University of 
Tasmania for the hospitality offered to him during his stay there in the 
summer of 2003. AB acknowledges SNI as well as CIC and CONACyT grants under 
projects 4.10 and 32395-E, respectively.

\begin{appendix}

\section{}

At the one loop level, the ${\cal A}$-term in Eq.~(\ref{eq:FEX}) corresponds 
to setting $F(k^2)=1$ and evaluating the integral~:
\begin{eqnarray}
 I \equiv  \int \frac{d^dk}{k^2 q^4} \; \left[(1-d) q^2 k \cdot p - 
2 \Delta^2  \right]  \;.
\end{eqnarray}
Re-writing $k \cdot p= (k^2+p^2-q^2)/2$ and 
$\Delta^2=(q^4 + k^4 + p^4 - 2 k^2 p^2 - 2 q^2 (k^2+p^2) )/4$, and using the
fact that the $d$-dimensional integration without the presence of external
momenta is zero, we simply arrive at the following expression~:
\begin{eqnarray}
     I = \frac{p^2}{2} \; \left[(3-d) \int \frac{d^dk}{k^2 q^2}
- p^2 \int \frac{d^dk}{k^2 q^4}   \right] \;.
\end{eqnarray}
Adopting the notation
\begin{eqnarray}
    J(\nu_1, \nu_2) &=& \int \frac{d^dk}{(k^2)^{\nu_1} (q^2)^{\nu_2}} \;,
\end{eqnarray}
we can now make use of the recursion relation,~\cite{davydychev2},
\begin{eqnarray}
\nonumber      J(\nu_1,\nu_2+1) &=& \frac{-1}{\nu_2 p^2} \; 
\Big[ (d-2 \nu_1 - \nu_2) J(\nu_1,\nu_2) - J(\nu_1-1,\nu_2+1)  \Big] \;,
\end{eqnarray}
to deduce that $I$ is identically zero in arbitrary dimensions. What
happens to all orders? We now have to evaluate the integral ${\cal I}$~:
\begin{eqnarray}
{\cal I} \equiv \frac{1}{2} \int \frac{d^dk}{k^2 q^4}  
\left[ 1 + \frac{F(k^2)}{F(p^2)}\right]
\left[(1-d) q^2 k \cdot p - 
2 \Delta^2  \right]  ,
\end{eqnarray}
which reduces to $I$ at the one loop level. We proceed by dividing the 
$d$-dimensional integral into radial and angular integrals so that
$d^dk = dk \, k^{d-1} {\rm sin}^{d-2} {\rm \theta} \, d{\rm \theta} \,
\Omega_{d-2}$, where $\Omega_{d-2}=2 \pi^{(d-1)/2}/ \Gamma((d-1)/2)$. 
The angular integral to be evaluated is~:
\begin{eqnarray}
\hspace{-15mm} 
\hat{I} &=&    \int_0^{\pi} d \theta  \frac{{\rm sin}^{d-2} \theta}{q^4}
 \Big [2 (d-2) k^2 p^2 {\rm cos}^2 \theta 
 + (1-d) k p (k^2+p^2)  {\rm cos} \theta + 2 k^2 p^2 \Big].   
\end{eqnarray}
Note that the integrand is proportional to the full derivative
$(d/d\theta)[{\rm sin}^{(d-1)}\theta / q^2 ]$. Thus the result is zero
because ${\rm sin}0={\rm sin}\pi=0$ at the integration limits.

\end{appendix}

\vfil\eject
\hsize=16.5cm

\vfil\eject
\vskip 1cm


\baselineskip=8mm
\begin{thebibliography}{999}
%
\bibitem{WGTI} J.C. Ward, Phys. Rev. {\bf 78} (1950); H.S. Green, Proc. Phys. 
Soc. (London) {\bf A66} 873 (1953); Y. Takahashi, Nuovo Cimento {\bf 6} 371 
(1957).
%
\bibitem{LKFT} L.D. Landau and I.M. Khalatnikov, Zh. Eksp. Teor. Fiz. {\bf 29} 
89 (1956); L.D. Landau and I.M. Khalatnikov, Sov. Phys. JETP {\bf 2} 69
(1956); E.S. Fradkin, Sov. Phys. JETP {\bf 2} 361 (1956).
%
\bibitem{zumino} K. Johnson and 
B. Zumino, Phys. Rev. Lett. {\bf 3} 351 (1959); B. Zumino, J. Math. Phys. 
{\bf 1} 1 (1960).
%
\bibitem{BC} J.S. Ball and T.W. Chiu, Phys. Rev. {\bf D22}, 2542 (1980).
%
\bibitem{King} J.M. Cornwall, Phys. Rev. {\bf D26}, 1453 (1983); 
J.E. King, Phys. Rev. {\bf D27}, 1821 (1983).
%
\bibitem{Brown} N. Brown and N. Dorey, Mod. Phys. Lett. {\bf A6}, 317 (1991).
%
\bibitem{CP} D.C. Curtis and M.R. Pennington, Phys. Rev. {\bf D42}, 4165 
(1990).
%
\bibitem{BP1} A. Bashir and M.R. Pennington, Phys. Rev. {\bf{D50}}, 7679 
(1994).
%
\bibitem{Ross} E.G. Floratos, D.A. Ross and C.T. Sachrajda, Nucl. Phys. 
{\bf B129}, 66 (1977).
%
\bibitem{Tarasov} O.V. Tarasov, JINR P2-82-900 (1982);
S.A. Larin, NIKHEF-H/92-18, hep-ph/9302240, in; Proc. Intern. Baksan 
School on Particles and Cosmology, eds. E.N. Alexeev, V.A. Matveev, Kh.S. 
Nirov and V.A. Rubakov (World Scientific, Singapore, 1994).
%
\bibitem{BKP1} A. Bashir, A. K{\i}z{\i}lers\"{u} and M.R. Pennington, 
Phys. Rev. {\bf D57} 1242 (1998).
%
\bibitem{BR} C.J. Burden and C.D. Roberts, Phys. Rev. {\bf D47} 5581 (1993).
%
\bibitem{Dong} Z. Dong, H.J. Munczek and C.D. Roberts, Phys. Lett.
{\bf 333}, 536 (1994).
%
\bibitem{BR1} A. Bashir and A. Raya, Phys. Rev. {\bf D64} 105001 (2001).
%
\bibitem{davydychev} A.I. Davydychev, P. Osland and L. Saks,
Phys. Rev. {\bf D63} 014022 (2001).
%
\bibitem{Ayse} A. K{\i}z{\i}lers\"u, M. Reenders and M.R. Pennington, Phys.
Rev. {\bf D52}, 1242 (1995).
%
\bibitem{Tarasov1} J. Fleischer, F. Jegerlehner, O.V. Tarasov and O.L. 
Veretin, Nucl. Phys. {\bf B539} 671 (1999); Erratum-ibid {\bf B571} 511
(2000).
%
\bibitem{BKP2} A. Bashir, A. K{\i}z{\i}lers\"{u} and M.R. Pennington, 
Phys. Rev. {\bf D62} 085002 (2000).
%
\bibitem{B1} A. Bashir, Phys. Lett. {\bf B491} 280 (2000).
%
\bibitem{davydychev1} A.I. Davydychev, Phys. Rev. {\bf D61} 087701 (2000).
%
\bibitem{DOT} A.I. Davydychev, P. Osland and O.V. Tarasov, Phys. 
Rev. {\bf D54} 4087 (1996).
%
\bibitem{RD03} R. Delbourgo, J. Phys. {\bf A36} 11697 (2003). 
%
\bibitem{davydychev2} F.A. Berends, A.I. Davydychev and V.A. Smirnov,
Nucl. Phys. {\bf B478} 59 (1996).
%
\end{thebibliography}
\end{document}